\documentclass[aps,prl,twocolumn,groupedaddress,showpacs]{revtex4}

\usepackage{graphicx}

\begin{document}


\title{On the possibility of an excitonic insulator at the
semiconductor-semimetal transition}


\author{Franz X. Bronold and Holger Fehske}
\affiliation{Institut f\"ur Physik, Ernst-Moritz-Arndt-Universit\"at Greifswald, D-17489 
Greifswald, Germany}


\date{\today}

\begin{abstract}
We calculate the critical temperature below which an excitonic insulator 
exists at the pressure-induced semiconductor-semimetal transition. Our 
approach is based on an effective-mass model for valence and conduction 
band electrons interacting via a statically screened Coulomb potential.
Assuming pressure to control the energy gap, we derive, in the spirit of
a BEC-BCS crossover scenario, a set of equations which determines,
as a function of the energy gap (pressure), the chemical potentials
for the two bands, the screening wave number, and the critical temperature. 
We (i) show that in leading order the chemical 
potentials are not affected by the exciton states, (ii) verify 
that on the strong coupling (semiconductor) side the critical temperatures 
obtained from the linearized gap equation coincide with the transition
temperatures for BEC of non-interacting bosons,
(iii) demonstrate that mass asymmetry strongly suppresses 
BCS-type pairing, and (iv) discuss in the context of our theory recent experimental 
claims for exciton condensation in ${\rm TmSe_{0.45}Te_{0.55}}$.
\end{abstract}

\pacs{71.35.-y, 71.30.+h, 71.28.+d, 75.20.Hr, 75.30.Mb}

\maketitle

\section{\label{Intro}I. Introduction}

More then four decades ago Mott~\cite{Mott61} realized that a semimetal (SM) 
with a low density of free charge carriers is unstable against an insulating, 
semiconducting (SC) state, because the Coulomb interaction binds 
conduction band electrons and valence band holes 
to excitons. At around the same time, Knox~\cite{Knox63} made the important 
observation that a SC whose energy gap is smaller then the exciton binding energy 
has to be unstable against the spontanous formation of excitons. Understood as  
a global instability, the pairing of electrons and holes is expected to lead 
to a macroscopic, phase coherent quantum state or condensate -- the excitonic 
insulator (EI) -- separating below a critical temperature the 
SC from the SM. Theoretically~\cite{KK64,desC65,KM65,KM66a,KM66b,JRK67,KK68,HR68,Kohn68}, 
the instability should occur in any material which can be tuned from an indirect gap 
semiconductor to a semimetal, for instance, by applying pressure, uniform stress, or 
by optical pumping. Yet, experimental efforts to establish the EI in real compounds 
largely failed.

It is only until recently that detailed studies of the pressure-induced
SC-SM transition in ${\rm TmSe_{0.45}Te_{0.55}}$ by Wachter and coworkers strongly 
suggested the existence of an EI~\cite{NW90,BSW91,W01,WBM04}. In particular, the 
anomalous increase of the electrical resistivity in a narrow pressure range around 
8~kbar indicated the appearance of a new phase below 250~K~\cite{NW90,BSW91}. 
Wachter and coworkers suggested that this new phase might be an ``excitonic insulator'',
and, assuming pressure to modify only the energy gap ${E_g}$, constructed a phase 
diagram for ${\rm TmSe_{0.45}Te_{0.55}}$ in the ${E_g-T}$ plane~\cite{BSW91}. Later 
they found in the same material a linear increase of the thermal diffusivity below
20~K and related this to a superfluid exciton state~\cite{WBM04}. Both excitonic 
phases are located on the SC side of the SC-SM transition in ${\rm TmSe_{0.45}Te_{0.55}}$.

In the present paper we perform the first theoretical analysis of these astonishing
experimental findings, staying strictly within the bounds of the concept of an
EI as a superfluid condensate of electron-hole pairs, similar to a condensate of 
Cooper pairs in a superconductor. (We exclusively use the acronym EI to denote this 
state and refer to condensation when electron-hole pairs enter this state.) In contrast 
to the early theoretical works~\cite{desC65,KK64,KM65,KM66a,KM66b,JRK67,KK68,HR68,Kohn68}, we (i) 
self-consistently 
calculate the phase boundary of the EI, (ii) analyze the EI in the spirit of a BEC-BCS 
crossover, and (iii) investigate also the halo of the EI, that is, 
the region surrounding the EI on the SC side.  

The creation of an exciton condensate is usually attempted
by optical pumping of suitable semiconductor structures~\cite{LEK04}.
So far, without success. The main obstacle is the far-off-equilibrium situation
caused by optical excitation. In contrast, pressure-induced generation of
excitons occurs at thermal equilibrium, which is much
more favorable for condensation. Semiconducting, pressure-sensitive mixed
valence materials, such as ${\rm TmSe_{0.45}Te_{0.55}}$, offer therefore a very
promising route towards exciton condensation. It is even conceivable to use
this class of materials for implementing recent proposals of coherent
transport across EI-SM junctions~\cite{RS05,WPX05}. The validation of Wachter
and coworkers' experiments would thus have a tremendous impact on the field
of exciton condensation.

The excitonic instability is driven by the Coulomb interaction leading to pairing of 
conduction band (CB) electrons with valence band (VB) holes. Of particular importance is 
therefore to self-consistently determine its weakening when the external parameter, for 
instance pressure, pushes 
the material from the SC side, with only a few thermally excited charge carriers 
available for screening, to the SM side, with a huge number of charge carriers.  
The strength of the Coulomb interaction determines on which side of the SC-SM 
transition the system is. As long as the Coulomb 
interaction is strong enough to support excitons, that is, as long as the
exciton binding energy is positive, the material is on the SC side. The 
vanishing of the binding energy (Mott effect~\cite{Mott61}) defines then the 
SC-SM transition.
\begin{figure}[t]
\includegraphics[width=0.90\linewidth]{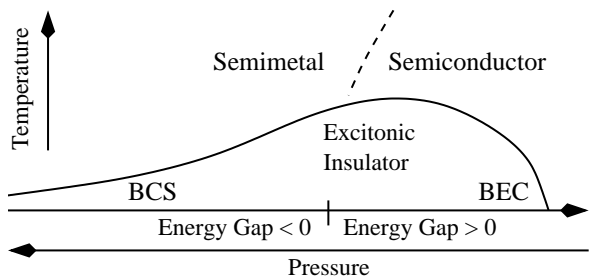}
\caption{\label{Cartoon}
Qualitative sketch of the phase boundary of an EI with equal band masses.
(Adopted from Ref.~\cite{Kohn68}.)}
\end{figure}

Depending from which side of the SC-SM transition the EI is approached
(see Fig. \ref{Cartoon}), the EI 
typifies either a Bose-Einstein (BE) condensate of tightly bound excitons (SC side) 
or a BCS condensate of loosely bound electron-hole pairs (SM side)~\cite{Leggett80,CN82}. 
A characteristic feature of BEC of excitons on the SC side is that 
formation of excitons and phase coherence (condensation)
occur at different temperatures~\cite{CN82,NSR85}. While excitons already exist 
below the temperature ${T_M}$ set by the Mott effect, the 
EI, understood as a genuine condensate of excitons, occurs only below the
critical temperature ${T_c < T_M}$. Thus, on the SC side, the EI is embedded in a 
region containing excitons in addition to VB holes and CB electrons. We call
this region halo. On the SM side, 
in contrast, Cooper-type electron-hole pairs form and condense at the same 
temperature ${T_c}$. The difference between an EI sitting, respectively, 
on the SC and SM side of the SC-SM transition has 
important consequences for the interpretation of the ${\rm TmSe_{0.45}Te_{0.55}}$
data.

After the seminal studies by Leggett~\cite{Leggett80}, Comte and Nozi\`eres~\cite{CN82}, 
and Nozi\`eres and Schmitt-Rink~\cite{NSR85}, the transition from BE to BCS condensation  
has been extensively studied in the past, mostly with an eye on short-coherence length
superconductors~\cite{DZ92,CG93,Haussmann93,MRE93,Roepke94,MJL99,PPS04}, but also 
electron-hole gases in semiconductors~\cite{CG88,SC99}, and, more recently, 
trapped atomic Fermi gases~\cite{OG02,CSTL05} have been analysed from this 
point of view. From these analytical and numerical investigations it is known
that the transition from BE to BCS condensation is smooth. It is also known 
that diagrammatic approaches usually capture the crossover only 
when the BCS equation for the order parameter is augmented by an equation for 
the chemical potential. At zero temperature, it suffices to calculate both quantities in
meanfield approximation~\cite{Leggett80,CN82}. At finite temperatures, however, the chemical 
potential has to be at least determined within a T-matrix approximation which accounts 
for the excitation of collective modes, i.e. in the context of an EI, of excitons 
with finite center-of-mass momentum~\cite{NSR85}.  

In order to discuss the EI at the pressure-induced SC-SM transition in terms of a BEC-BCS 
crossover, we need therefore not only equations for the order parameter and
the screening wave number (to account for the Mott effect) but also for the CB electron 
and VB  hole chemical potentials. Because pressure controls the chemical potentials, 
which in turn are constrained by charge neutrality, the meanfield approximation turns
out, even for finite temperatures, to be sufficient on the (weak coupling) SM side and 
the (strong coupling) SC side, in contrast to what one would expect from related 
diagrammatic studies of the crossover problem in high-${T_c}$ materials and atomic Fermi
gases~\cite{DZ92,Haussmann93,MRE93,MJL99,PPS04,OG02,CSTL05}. Using 
the Thouless criterion~\cite{Thouless60} (see also~\cite{KK68}) to determine the 
transition temperature for BEC directly from the normal phase electron-hole T-matrix, 
we verify that the transition temperatures obtained from the meanfield 
equations indeed coincide on the SC side with the transition temperatures for BEC of
non-interacting bosonic excitons. 
From the normal phase electron-hole T-matrix we furthermore extract the 
temperature ${T_M}$ above which excitons cease to exist and construct a 
mass-action-law to determine the composition of the EI's halo.

As far as Wachter and coworkers' experiments~\cite{NW90,BSW91,W01,WBM04} are concerned
we come to the following conclusions. The phase boundary constructed from the resistivity
data does not embrace the EI.
Instead, it most probably reflects that part of the halo of the EI where excitons prevail 
over free electrons and holes and give rise to an efficient additional scattering channel
which leads to the observed resistivity anomaly. More remarkable is however that from our
theoretical results we would expect the EI -- that is, the macroscopic, phase coherent 
condensate -- exactly in the temperature range where the linear increase of the thermal 
diffusivity was observed~\cite{WBM04}. In analogy to liquid He~4, Wachter and coworkers
suggest that the increase could be due to the second sound of a superfluid exciton
liquid. We, on the other hand, find the EI entirely on the SC side, where, within our
approximations, it constitutes a BEC of ideal bosonic excitons. The BCS side of the EI,
where it is unclear whether superfluidity can be realized~\cite{KM66b} or 
not~\cite{Zittartz68b}, is
strongly suppressed for finite temperatures because of the large asymmetry between the CB 
and VB masses in ${\rm TmSe_{0.45}Te_{0.55}}$. Of course, an ideal gas of bosons cannot 
become superfluid and thus cannot feature a second sound. In reality, however, excitons
interact and could, when the interaction is dominantly repulsive, give rise to it.
Thus, Wachter and coworkers~\cite{WBM04} may have seen 
an exciton condensate.

In the next section we first introduce an effective-mass model on which our 
calculation is based and then give a description of our calculational 
scheme. In section III we 
present numerical results for the phase boundary ${T_c(E_g)}$ of the EI, 
first for equal band masses, where we obtain a steeple-like ${T_c(E_g)}$,
directly reflecting the different condensation mechanisms on the SC and SM side, 
and then for asymmetric band masses, applicable to Tm[Se,Te] compounds, where we 
find a strongly suppressed 
${T_c(E_g)}$ on the SM side and thus an EI sitting almost entirely on the SC
side of the SC-SM transition.  
We then relate in section IV our results to the experimental data for the
pressure induced SC-SM transition in ${\rm TmSe_{0.45}Te_{0.55}}$ and 
conclude in section V.

\section{\label{Formalism}II. Formalism}
\subsection{A. Model}
The excitonic instability arises because of the Coulomb attraction between 
electrons in the lowest CB ($i=1$) and holes in the highest VB ($i=2$), with 
an indirect energy gap separating the two bands. Since the spin algebra would 
unnecessarily mask the many-body theoretical concepts we want to discuss,
we focus on spinless fermions. Only in  
section III, where we make contact with experiments, we include the spin in
our calculation.

Keeping only the dominant term 
of the Coulomb interaction, which we assume to be statically screened, 
an effective-mass model for studying the (spinless) Wannier-type EI is 
\begin{equation}
H=\sum_{{\bf k},i}{e}_i({\bf k})
c_{i,{\bf k}}^\dagger c_{i,{\bf k}}
+\frac{1}{2}\sum_{{\bf q}} V_s({\bf q}) \rho({\bf q})\rho(-{\bf q}),
\label{model}
\end{equation}
with $\rho({\bf q})=\sum_{i,{\bf k}} c_{i,{\bf k}+{\bf q}}^\dagger c_{i,{\bf k}}$
the total charge density and $V_s({\bf q})=(4\pi e^2/\epsilon_0)/(q^2+q_s^2)$. 
The screening wave number $q_s$ depends
on the CB electron and VB hole density and will be determined self-consistently
in the course of the calculation; $\epsilon_0$ is the background dielectric constant.
The momenta ${\bf k}$ for CB electrons and VB holes are measured
from the respective extrema of the bands which are separated by half of a
reciprocal lattice vector ${\bf w}={\bf K}/2$. 
We consider the case where both the VB and the CB band have one extremum.
Assuming isotropic 
effective masses $m_i$, the band dispersions read ($\hbar=1$ throughout the paper)
\begin{equation}
{e}_1({\bf k})={E_g}+\varepsilon_1({\bf k})-\mu ,~~\\
{e}_2({\bf k})=-\varepsilon_2({\bf k})-\mu -\Sigma_0({\bf k}),
\label{dispersion}
\end{equation}
with $\varepsilon_i({\bf k})={\bf k}^2/2m_i$, ${E_g}$ the bare energy gap 
which can be varied continuously through zero 
under pressure, and $\mu$ the chemical potential. The band structure (\ref{dispersion})
refers to the unexcited crystal at ${T=0}$ with an empty CB and a full VB whose 
selfenergy $\Sigma_0({\bf k})$ has to be subtracted~\cite{Zimmermann76}.

For model (\ref{model}) to be applicable for the description of an EI we have 
to assume that through the SC-SM transition all parameters, except the energy gap
${E_g}$ and the screening wave number $q_s$, vary weakly and can be kept constant. 
Since model (\ref{model}) treats an indirect gap semiconductor 
as a direct gap semiconductor, effects due to the indirect gap~\cite{HR68}, in 
particular, multi-valley effects (when the valleys are not identical) and the balancing 
of the finite momentum ${\bf w}$ of the electron-hole pairs by the lattice leading, 
for instance, to (small) density modulations in the EI phase are beyond the scope 
of the present paper. 

\subsection{\label{MFA}B. Meanfield approximation}

In formal analogy to the theory of superconductivity, we employ matrix propagators 
(in the band indices $i=1,2$) 
\begin{equation}
{G}_{ij}({\bf k},\tau)=-\langle {T}_\tau c_{{\bf k},i}(\tau) 
c_{{\bf k},j}^\dagger(0) \rangle, 
\end{equation}
with $0\leq \tau \leq \beta$ and ${T}_\tau[...]$ the time-ordering operator 
with respect to $\tau$. The diagonal elements ${G}_{ii}$ denote the (normal) propagators 
for electrons in the CB and VB, respectively, whereas the off-diagonal (anomalous) 
propagators describe the phase coherence of electron-hole pairs. Finite anomalous 
propagators signal the condensate and thus the EI.

In Matsubara space, the matrix propagator satisfies a Dyson equation, 
\begin{equation}
{G}(k)={g}(k)+{g}(k)\Sigma(k){G}(k),
\label{Dyson}
\end{equation}
with a diagonal matrix 
\begin{eqnarray}
{g}_{ij}(k)=[i\omega_n-{e}_i({\bf k}]^{-1}\delta_{ij},
\label{bareGF}
\end{eqnarray} 
denoting the bare propagator and a 2$\times$2 selfenergy matrix $\Sigma$,
which we assume to be given by a skeleton expansion, that is, 
in terms of the fully dressed propagator ${G}$. The variable 
$k=({\bf k},i\omega_n)$ denotes a four-vector with ${\bf k}$ a momentum 
and $\omega_n=(2n+1)\pi/\beta$ ($n$ integer) a fermionic Matsubara frequency.

Separating the selfenergy into diagonal (normal) and off-diagonal
(anomalous) parts and suppressing the $k$ dependence, we rewrite the Dyson 
equation as 
\begin{eqnarray}
{G}&=&{\cal G}+{\cal G}\Sigma^{A}{G},
\label{Dyson_a}\\
{\cal G}&=&{g}+{g}\Sigma^{N}{\cal G}
\label{Dyson_b}
\end{eqnarray}
with $\Sigma^{A}$ and $\Sigma^{N}$ the anomalous and normal 
selfenergy, respectively. 

The separation of the Dyson equation is particularly useful for the
calculation of the transition temperature ${T_c}$.  
We first note from Eq.~(\ref{Dyson_b}) that ${\cal G}$ is diagonal. 
Thus, Eq.~(\ref{Dyson_a}) reduces to 
\begin{eqnarray}
{G}^{-1}=
\bigg(
\begin{array}{cc}
{\cal G}_{11}^{-1}-{\cal G}_{22} |\Sigma^{A}_{12}|^2 &
\big[{\cal G}_{11} {G}_{22} \Sigma^{A}_{12}\big]^{-1} \\
\big[{\cal G}_{22} {G}_{11} \Sigma^{A}_{21}\big]^{-1} &
{\cal G}_{22}^{-1}-{\cal G}_{11} |\Sigma^{A}_{12}|^2
\end{array}
\bigg).
\label{D1}
\end{eqnarray}
In the vicinity of the transition temperature ${T_c}$ the 
anomalous selfenergy becomes vanishingly small and Eq.~(\ref{D1}) 
can be linearized with respect to $\Sigma^{A}$. The anomalous 
propagator ${G}_{12}$ is then given by 
\begin{eqnarray}
{G}_{12}&=&{\cal G}_{11} {\cal G}_{22} \Sigma^{A}_{12},\label{DD}
\end{eqnarray}
where the propagators ${\cal G}_{ii}$ satisfy Eq.~(\ref{Dyson_b}) with 
$\Sigma^{N}$ calculated from ${\cal G}_{ii}$ alone; ${G}_{12}$ and 
${G}_{21}$ do not enter $\Sigma^{N}$ anymore. Thus, within the linearized
theory, the calculation of the normal propagators is decoupled from the 
calculation of the anomalous propagators. From Eq.~(\ref{DD}), we see
moreover that ${G}_{12}\neq 0$ is equivalent to $\Sigma^{A}_{12}\neq 0$.   
Accordingly, the anomalous selfenergy can be considered as an order parameter.
\begin{figure}[t]
\includegraphics[width=0.90\linewidth]{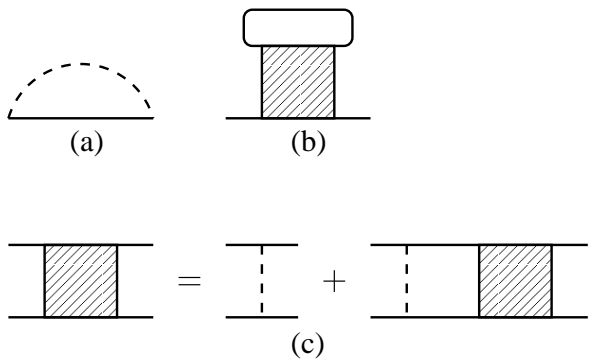}
\caption{\label{Fig1} (a) exchange selfenergy $\Sigma^X$ to be used in the 
meanfield approximation, (b) ladder selfenergy $\Sigma^L$ additionally 
used in the T-matrix approximation, and (c) ladder equation for the T-matrix.
Solid (dashed) lines denote matrix propagators (statically screened
Coulomb potentials).} 
\end{figure}

To proceed we have to specify the selfenergies $\Sigma^{N}$ and 
$\Sigma^{A}$. In the meanfield approximation both are given 
by the exchange energy $\Sigma^{X}$ shown in Fig.~\ref{Fig1}a. Applying standard 
diagrammatic rules~\cite{Schrieffer}, we find 
\begin{eqnarray}
\Sigma^{X}_{ii}({\bf k})=-\int\frac{d{\bf k'}}{(2\pi)^3}
V_s({\bf k-k'}) n_{\rm F}(\bar{e}_i({\bf k'})\label{SN}
\end{eqnarray}
for the normal selfenergies and  
\begin{eqnarray}
\Delta({\bf k})=\int\frac{d{\bf k'}}{(2\pi)^3} 
V_s({\bf k-k'})Q({\bf k'})\Delta({\bf k'}) \label{SA}
\end{eqnarray}
with 
\begin{eqnarray}
Q({\bf k})= \frac{n_{\rm F}(\bar{e}_2({\bf k}))-n_{\rm F}
(\bar{e}_1({\bf k}))}{\bar{e}_1({\bf k})-\bar{e}_2({\bf k})}
\label{Qgap}
\end{eqnarray}
for the anomalous selfenergy $\Delta({\bf k})=\Sigma^{X}_{12}({\bf k})$,
where $n_{\rm F}$ denotes the Fermi function. 

Note, the normal selfenergies contain no Coulomb-hole term because 
we replaced directly in the Hamiltonian the bare Coulomb interaction by the 
statically screened one (static screening approximation); the correlation
energy is thus partly suppressed. As a consequence, excitonic phases are  
energetically more favorable than electron-hole liquid phases~\cite{R77}, 
independent of the mass ratio and the number of valleys, and in contrast to,
for instance, what one finds in the quasi-static approximation~\cite{BFR06}, 
which performs the replacement (with a modified screened Coulomb potential) after 
the selfenergies have been calculated in the random-phase approximation~\cite{HS84}. 
Neither approximation, however, should be used to study the competition 
between the two phases, because both do not give a reliable correlation energy 
for intermediate densities, that is, close to the SC-SM transition, where 
excitons should be included in the correlation energy~\cite{RD79}. In addition, 
scattering processes beyond our model, for instance, between valleys or with 
phonons, also affect the competition between excitonic and electron-hole liquid 
phases. An investigation of the competition between the two is thus a complex
task, beyond the scope of the present paper, which simply assumes excitons to 
be favored from the start. The static screening model is thus sufficient.

The single particle energies entering Eqs.~(\ref{SN}) and (\ref{Qgap}) are the 
self-consistent solutions of
\begin{eqnarray}
{\bar{e}_i}({\bf k})={e}_i({\bf k})+\Sigma^{X}_{ii}({\bf k}).
\label{renDis}
\end{eqnarray}
The main effect of the exchange energy on the particle dispersions is a rigid,
${\bf k}$-independent energy shift which we incorporate into chemical 
potentials for the CB electrons and VB holes. Writing the renormalized 
dispersions as 
\begin{eqnarray}
{\bar{e}_i}({\bf k})=(-1)^{i+1}(\varepsilon_i({\bf k})-\mu_i)
\label{rendis}
\end{eqnarray}
yields for the electron and hole chemical potentials 
\begin{eqnarray}
\mu_1&=&\mu-{E_g}-\Delta {e}_1,\label{mu1}\\
\mu_2&=&\Delta {e}_2-\mu\label{mu2},
\end{eqnarray}
with energy shifts satisfying 
\begin{eqnarray}
\Delta{e}_i=(-1)^i \int\frac{d{\bf k}}{(2\pi)^3}
V_s({\bf k}) n_{\rm F}(\varepsilon_i({\bf k})-\mu_i)\label{Delta1}.
\end{eqnarray}
To derive Eq.~({\ref{Delta1}) for $i=2$, the selfenergy $\Sigma_0$ of the full VB has 
to be calculated within the same approximation and then drops out.
Together with Eq.~(\ref{Delta1}), Eqs.~(\ref{mu1}) and (\ref{mu2}) effectively define an 
electron-hole representation which is convenient for the calculation of the screening 
parameter $q_s$.

In order to obtain a closed set of equations for the three unknown parameters
$\mu$, $\mu_1$, and $\mu_2$, we augment Eqs.~(\ref{mu1}) and (\ref{mu2})
by the condition of charge neutrality which forces CB electron and VB hole 
densities, given in the meanfield approximation, respectively, by 
\begin{eqnarray}
n_1&=& \int\frac{d{\bf k}}{(2\pi)^3}n_{\rm F}(\varepsilon_1({\bf k})-\mu_1),\label{n1}\\
\bar{n}_2&=&\int\frac{d{\bf k}}{(2\pi)^3}n_{\rm F}(\varepsilon_2({\bf k})-\mu_2)\label{n2},
\end{eqnarray}
to be equal. Thus, charge neutrality leads to the constraint
\begin{eqnarray}
\int\frac{d{\bf k}}{(2\pi)^3}\left[ n_{\rm F}(\varepsilon_1({\bf k})-\mu_1)
- n_{\rm F}(\varepsilon_2({\bf k})-\mu_2)\right]=0,\label{neutral}
\end{eqnarray}
which we use to eliminate $\mu$ in Eqs.~(\ref{mu1}) and (\ref{mu2}). 
For that purpose, we add Eqs.~(\ref{mu1}) and (\ref{mu2}) 
\begin{eqnarray}
\mu_1+\mu_2=-{E_g}+\sum_i\int\frac{d{\bf k}}{(2\pi)^3}V_s({\bf k})
n_{\rm F}(\varepsilon_i({\bf k})-\mu_i)
\label{Egren}
\end{eqnarray}
and combine this equation with the charge neutrality constraint (\ref{neutral}).
Physically, the sum of the CB electron and VB hole chemical potentials
is the negative of the renormalized energy gap:
\begin{eqnarray}
\bar{E}_{g}=\bar{e}_1(0)-\bar{e}_2(0)=-\mu_1-\mu_2. 
\label{Egbar}
\end{eqnarray}

With the individual chemical potentials for each species at our disposal, the 
screening parameter $q_s$ is given by 
\begin{eqnarray}
q_s^2=\frac{4\pi e^2}{\epsilon_0}\left( \frac{\partial}{\partial\mu_1} n_1 
+ \frac{\partial}{\partial\mu_2} \bar{n}_2\right).
\label{screening}
\end{eqnarray}

Once $\mu_1, \mu_2$, and $q_s^2$ are calculated, the kernel of the 
gap equation (\ref{SA}) is known in the whole ${E_g-T}$ plane. 
For a given energy gap ${E_g}$, we can thus determine the transition 
temperature ${T_c}$ as the highest temperature for which the homogeneous 
integral equation (\ref{SA}) has a nontrivial solution $\Delta({\bf k})\neq 0$. 

\subsection{C. T-matrix approximation}
The meanfield theory presented in the previous subsection determines 
${T_c(E_g})$ on the strong coupling (SC) as well as on 
the weak coupling (SM) side. The transition temperatures on the SC 
side turn out to coincide with the BEC transition temperatures
for a non-interacting boson gas of excitons (see section III), 
as it should~\cite{KK68,NSR85}. From the 
perspective of diagrammatic BEC-BCS crossover 
theories~\cite{NSR85,DZ92,Haussmann93,MRE93,MJL99,PPS04,OG02,CSTL05} 
this is however surprising 
because the meanfield theory does not account for excitons existing 
on the SC side already above the transition temperature ${T_c(E_g})$. 
These excitons are expected to at least modify the chemical potentials
$\mu_i$. However, the charge neutrality constraint, inherent in
a pressure-driven BEC-BCS crossover, leads to a cancellation of the 
leading order corrections to the chemical potentials.

Diagrammatic BEC-BCS crossover theories~\cite{NSR85,DZ92,Haussmann93,MRE93,MJL99,PPS04,OG02,CSTL05}
usually leave the gap equation~(\ref{SA}) unchanged but calculate the charge densities 
from normal propagators dressed not only with the exchange selfenergy $\Sigma^X$ 
(Fig. \ref{Fig1}a) but also with the ladder selfenergy $\Sigma^L$ (Fig. \ref{Fig1}b),
which takes, via the electron-hole T-matrix (Fig. \ref{Fig1}c),  
normal phase excitons into account. Expanding the spectral functions
to which the total selfenergy gives rise to with respect to the imaginary part 
of $\Sigma^L$~\cite{ZS85} (see appendix A), 
the CB electron (VB hole) density can be written as $n_1=n_1^{f}+n_1^{b}$ 
($\bar{n}_2=\bar{n}_2^{f}+\bar{n}_2^{b}$) with the superscripts `f' and `b' 
denoting, respectively, the part of the total density coming from free electrons 
(holes), with a modified dispersion but a spectral weight unity, and 
the part arising from bound electron-hole pairs.
The charge neutrality condition (\ref{neutral}) becomes therefore
$n_1^{f}+n_1^{b}=\bar{n}_2^{f}+\bar{n}_2^{b}$. 
Now it is important to realize (or to verify by calculation, see appendix A) that
whenever a CB electron participates in an exciton, a VB hole has to 
do the same. Thus, $n_1^{b}=\bar{n}_2^{b}$, and the charge neutrality condition 
reduces to $n_1^{f}=\bar{n}_2^{f}$ which, except for the modifications
of the dispersions due to $\Sigma^L$ has the same form as in the meanfield
approximation. 

The above argument is independent from the approximation used to obtain
the electron-hole T-matrix. It is only based on charge neutrality 
and the fact that an exciton has to contain both an electron and a hole. The 
calculation shows that it is crucial to employ the skeleton expansion. 
As a result, the single particle dispersions are modified by $\Sigma^L$ and 
thus by excitons (see Eq.~(\ref{eladder})). The ladder approximation 
(Fig.~\ref{Fig1}c) we use to calculate the 
T-matrix does however not consistently account for feedback effects of this
kind; for that purpose, additional diagrams~\cite{Roepke94} 
or a systematic mode-coupling approach~\cite{CG88,CG93} should be implemented.
We consider therefore feedback of excitons to single particle states as higher order 
effects and ignore all of them, including the one already present in the 
ladder approximation. The charge neutrality condition in the ladder 
approximation reduces then to the meanfield charge neutrality condition. 
Note, feedback effects induce interactions between pairs~\cite{PPS04}.
Neglecting them implies therefore to stay within the framework of a
non-interacting, ideal gas of excitons. How to extract from the skeleton 
expansion 
(and not from an algebraic mapping which is 
questionable in the vicinity of the SC-SM transition) 
the interaction between electron-hole pairs, which not only 
contains the repulsive core at short distances because of the Pauli 
exclusion but also the weakly attractive tail at large distances, is to the 
best of our knowledge an unsolved problem and beyond the scope of the present 
paper. 

In appendix A we calculate the charge densities within the ladder approximation
shown in Fig. \ref{Fig1} using the separable approximation for the normal phase 
electron-hole T-matrix described in appendix B. Thereby, we proof by calculation 
that $n_1=\bar{n}_2$ reduces to $n_1^{f}=\bar{n}_2^{f}$. Because the screening 
parameter $q_s$, on the other hand, accounts only for screening due to free charge 
carriers it has to be calculated from $n_1^{f}$ and $\bar{n}_2^{f}$. 
Thus, even for finite temperatures, 
the meanfield approximation of the previous subsection is on par with T-matrix based 
BEC-BCS crossover theories~\cite{NSR85,DZ92,Haussmann93,MRE93,MJL99,PPS04,OG02,CSTL05}.

As an additional check, we now deduce the transition temperature on the 
SC side directly from the normal phase electron-hole T-matrix (which takes excitons
with finite center-of-mass momentum into account) using the Thouless 
criterion~\cite{Thouless60} (see also~\cite{KK68}) and verify that the
${T_c(E_g)}$ calculated within the meanfield approximation indeed coincides with the
transition temperatures obtained from the T-matrix. 
Thereby, we also verify that the neglect of the feedback of 
excitons on single particle states leads to a non-interacting bosonic gas of excitons.


\begin{widetext}
In the approximation depicted in Fig.~\ref{Fig1}c, the normal phase electron-hole
T-matrix is given by 
\begin{eqnarray}
\Lambda_{12}({\bf k},{\bf k'};{\bf q},i\Omega_n)=
V_s({\bf k-k'})+\int\frac{d{\bf k''}}{(2\pi)^3}V_s({\bf k-k''})
Q({\bf k''};{\bf q},i\Omega_n)\Lambda_{12}({\bf k''},{\bf k'};{\bf q},i\Omega_n)
\label{LadderEq}
\end{eqnarray}
with an electron-hole pair propagator 
\begin{eqnarray}
Q({\bf k};{\bf q},i\Omega_n)=
\frac{1-n_{\rm F}(\varepsilon_1({\bf k})-\mu_1)-n_{\rm F}(\varepsilon_2({\bf k}-{\bf q})-\mu_2)}
{\bar{E}_{g}+\varepsilon_1({\bf k})+\varepsilon_2({\bf k}-{\bf q})-i\Omega_n}.
\end{eqnarray}
Here, we used Eq.~(\ref{rendis}) to express renormalized dispersions $\bar{e}_i$ 
in terms of chemical potentials $\mu_i$ and definition (\ref{Egbar}) to introduce the 
renormalized energy gap; $\Omega_n=2n\pi/\beta$ ($n$ integer) are bosonic Matsubara 
frequencies and ${\bf q}$ denotes the center-of-mass momentum of an electron-hole
pair (and thus of an exciton). 

Taking advantage of the fact that the most important part of the normal phase electron-hole
T-matrix originates form the exciton state, we calculate in appendix B the T-matrix 
within a separable approximation. As a result, we obtain
\begin{eqnarray}
\Lambda_{12}({\bf k},{\bf k'};{\bf q},i\Omega_n)=
g(|{\bf k}-\frac{m_1}{M}{\bf q}|)\cdot 
D_X({\bf q},i\Omega_n-\bar{E}_{g}-\frac{q^2}{2M})
\cdot g(|{\bf k'}-\frac{m_1}{M}{\bf q}|)
\label{vertex}
\end{eqnarray}
with a renormalized exciton propagator
\begin{eqnarray}
D_X({\bf q},i\Omega_n)=\frac{-1}{i\Omega_n+B+M_X({\bf q},i\Omega_n)}
\label{Xprop}
\end{eqnarray}
and form factors 
\begin{eqnarray}
g(p)=\int_0^\infty \frac{dp'p'^2}{4\pi^2}V_s(p,p')\chi(p')=
\bigg[\frac{p^2}{2m}+B\bigg]\chi(p)
\label{Wannier}
\end{eqnarray}
defined in terms of a screened exciton wavefunction $\chi$ and a screened exciton 
binding energy $B$ (see below and appendix B).
Here, $V_s(p,p')$ denotes the screened Coulomb potential averaged over the angle 
between ${\bf p}$ and ${\bf p'}$, $M=m_1+m_2$ is the total mass, and $m=m_1 m_2/M$ 
is the reduced mass of an electron-hole pair.

The exciton propagator~(\ref{Xprop}) describes an exciton in a medium consisting of a finite 
density of CB electrons and VB holes. The screened exciton binding energy $B$ already 
accounts for the screening due to spectator particles not participating in the exciton. 
The spectators' Pauli blocking, on the other hand, is included through the selfenergy
\begin{eqnarray}
M_X({\bf q},i\Omega_n)=-\int \frac{d{\bf p}}{(2\pi)^3}\frac{[\frac{p^2}{2m}+B]^2}
{\frac{p^2}{2m}-i\Omega_n}\chi^2(p)
\bigg[n_{\rm F}(\varepsilon_1({\bf p}+\frac{m_1}{M}{\bf q})-\mu_1)
+n_{\rm F}(\varepsilon_2({\bf p}-\frac{m_2}{M}{\bf q})-\mu_2)\bigg].
\label{M}
\end{eqnarray}

To avoid unnecessary numerical work, we replace in Eq.~(\ref{Wannier}) 
the statically screened Coulomb potential by the Hulthen potential~\cite{HaugKoch},
\begin{eqnarray}
V_s(r)=\frac{e^2}{\epsilon_0}\frac{\exp{(-q_s r)}}{r}
\rightarrow 
V_H(r)=\frac{e^2}{\epsilon_0} \frac{2q_s}{\exp{(2q_sr)}-1},
\label{Hulthen}
\end{eqnarray}
which is a good approximation as long as the screening wave number $q_s$ is not 
too large. Indeed, in the limit $q_s\rightarrow 0$, the Hulthen potential reduces to the 
unscreened Coulomb potential. With the replacement (\ref{Hulthen}), Eq.~(\ref{Wannier}) can be 
solved analytically~\cite{HaugKoch}. 
The screened exciton wavefunction and the screened exciton binding energy for $a_X\cdot q_s<1$ are,
respectively, given by
\begin{eqnarray}
\chi(p)&=&
\frac{8\sqrt{\pi a_X^3(1-(a_X\cdot q_s)^2)}}{[(1-a_X\cdot q_s)^2+(a_X\cdot p)^2]
\cdot [(1+a_X\cdot q_s)^2+(a_X\cdot p)^2]},\label{chi}\\\nonumber\\
B&=&R_X(1-a_X\cdot q_s)^2,\label{Bbare}
\end{eqnarray}
\end{widetext}
where $R_X=1/2ma_X^2$ is the exciton Rydberg and 
$a_X=\epsilon_0/me^2$ is the exciton Bohr radius ($e$ bare electron charge). 
For $a_X\cdot q_s > 1$,
the  Hulthen potential does not support a bound state and we have to set 
$\chi=B=0$ in that parameter range. Notice, in the limit $q_s\rightarrow 0$, 
Eqs.~(\ref{chi}) and (\ref{Bbare}), respectively, reduce to the wave function and 
binding energy of an isolated exciton (no medium effects) as obtained from 
Eq.~(\ref{Wannier}) with $V_s$ replaced by the angle averaged unscreened Coulomb 
potential. 

The pole of the exciton propagator (\ref{Xprop}) determines the analytical structure
of the T-matrix. Physically, it gives the exciton
binding energy $\bar{B}({\bf q})$ renormalized by screening and phase space filling.
Assuming a weak ${\bf q}$-dependence, $\bar{B}({\bf q})\approx \bar{B}(0)\equiv\bar{B}$,
with the renormalized exciton binding energy $\bar{B}$ determined from
\begin{eqnarray}
\bar{B}=B+{\rm Re}M_X(0,i\Omega_n\rightarrow -\bar{B}+i\eta).
\label{SelfX}
\end{eqnarray}
In terms of $\bar{B}$ the exciton propagator can be rewritten as 
\begin{eqnarray}
D_X({\bf q},i\Omega_n)=\frac{-Z_X}{i\Omega_n+\bar{B}}
\label{Xprop1}
\end{eqnarray}
with an exciton spectral weight defined by 
\begin{eqnarray}
Z_X=1-\frac{\partial}{\partial\Omega}{\rm Re}M_X(0,\Omega+i\eta)|_{\Omega=-\bar{B}}.
\label{Zx}
\end{eqnarray}

The Thouless criterion states that BEC of excitons occurs when the normal phase electron-hole 
T-matrix diverges at ${\bf q}=0$ and $i\Omega_n=0$. Using Eq.~(\ref{vertex}) this 
implies $D_X^{-1}(0,-\bar{E}_{g})=0$. With Eq.~(\ref{Xprop1}), the transition temperature
${T_c}({E_g})$ is thus given by
\begin{eqnarray}
\bar{B}({E}_{g},{T_c})=\bar{E}_{g}({E}_{g},{T_c}),
\label{Thcrit}
\end{eqnarray}
where we displayed the dependence of the renormalized binding energy and
the renormalized energy gap on the bare energy gap and the temperature. 

In order to make 
the physical content of Eq.~(\ref{Thcrit}) more transparent, we rewrite it in terms of
chemical potentials. Recalling the definition (\ref{Egbar}) of the renormalized
energy gap, Eq.~(\ref{Thcrit}) becomes 
\begin{eqnarray}
\mu_X=\mu_1+\mu_2=-\bar{B},
\label{IdealBG}
\end{eqnarray}
which is equivalent to the criterion for BEC of a non-interacting Bose gas:
The transition occurs when the chemical potential $\mu_X$ of the bosons reaches 
the bottom of the boson band leading to a macroscopic occupation of the ${\bf q}=0$ state.

The Thouless criterion per se~\cite{Thouless60} can be used to determine 
${T_c}({E_g})$ on the SC~\cite{KK68} and the SM~\cite{KM65} side.
However, by construction, Eq.~(\ref{vertex}) describes the normal phase
electron-hole T-matrix only on the SC side, where excitons exist, and the separable 
approximation for the T-matrix is applicable. Thus, we can use the Thouless criterion 
only on the SC side.

The exciton binding energy obtained from Eq.~(\ref{SelfX}) enables us also to determine 
the phase boundary ${T_M}({E_g})$ between the SC and the SM and thus the halo of the 
EI, that is, the region between ${T_M}({E_g})$ and ${T_c}({E_g})$ on the SC side, where 
excitons, CB electrons and VB holes coexist. As originally 
suggested by Mott~\cite{Mott61}, we use the vanishing of $\bar{B}$ (Mott effect)
as a criterion for the SC-SM transition. Accordingly, ${T_M}({E_g})$ is given by 
\begin{eqnarray}
\bar{B}({E_g},{T_M})=0.
\label{Mcrit}
\end{eqnarray}

\section{\label{Results}III. Results}

To construct the phase boundary of the EI, we discretize the gap equation 
(\ref{SA}), and determine, for a given energy gap ${E_g}$, the temperature 
${T_c}({E_g})$ 
for which the determinant of the coefficient matrix of the resulting linear set of 
equations vanishes. For each pair (${E_g},T$) we supply the chemical potentials 
$\mu_1$ and $\mu_2$ together 
with the screening wave number $q_s$ by finding the simultaneous 
roots of Eqs.~(\ref{neutral}), (\ref{Egren}), and (\ref{screening}). 
To obtain the SC-SM phase boundary, we first determine, from  
Eq.~(\ref{SelfX}), the renormalized exciton binding energy $\bar{B}$ as a function of 
${E_g}$ and $T$, again providing for each pair (${E_g},T$) the chemical potentials 
and the screening wave number. After that, we search for a given ${E_g}$ for the 
temperature ${T_M}({E_g})$ which satisfies 
Eq.~(\ref{Mcrit}). In a similar way, we determine on the SC side, as a cross-check of 
the results obtained from the linearized gap equation, the critical temperature 
${T_c}({E_g})$ from the Thouless criterion (\ref{Thcrit}). 

We consider an isotropic system, angles can be thus integrated
analytically. All integrals over the magnitude of the momentum are done by Gaussian
integration except the $|{\bf k}|$-integration appearing in the gap equation, which 
we perform by
product integration, with the angle averaged Coulomb potential $V_s(k,k')$ as the weight
function. The logarithmic singularity of the kernel at $k=k'$, which arises on the SC side
for ${T}\rightarrow 0$, can thus be handled without problems. Recall, the 
separable approximation for the electron-hole T-matrix, on which the calculation of  
the renormalized exciton binding energy $\bar{B}$ is based, uses the Hulthen potential 
instead of the statically screened Coulomb potential. All results are presented
in exciton units, measuring energies and temperatures in exciton Rydbergs $R_X$ 
and lengths in exciton Bohr radii $a_X$. 
\begin{figure}[t]
\includegraphics[width=0.90\linewidth]{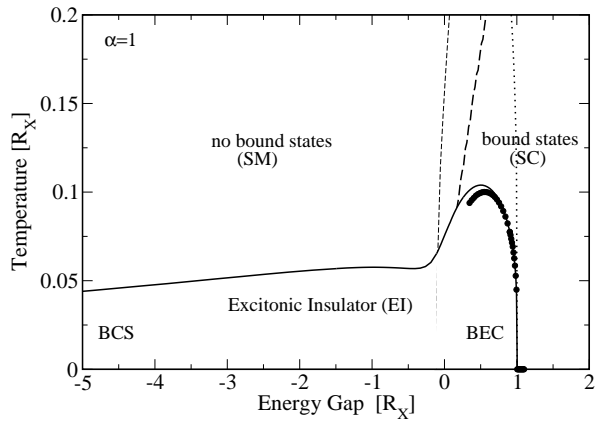}
\caption{\label{Fig2} Phase boundary of an EI with equal band
masses ($\alpha=m_1/m_2=1)$. The solid line (full circle) denotes
the critical temperature obtained from the linearized gap
equation (Thouless criterion for BEC). The dotted line shows
part of the phase boundary on the SC side when no
screening due to thermally excited charge carriers is taken into
account. The thick (thin) dashed line indicates the SC-SM transition,
defined by the vanishing of the renormalized (screened) binding energy,
taking screening and Pauli blocking (only screening) into account.}
\end{figure}

The phase boundary ${T_c}({E_g})$ for an EI with equal band masses 
($\alpha=m_1/m_2=1$) is presented in Fig.~\ref{Fig2}. The solid line shows
${T_c(E_g)}$ obtained from the linearized gap equation (\ref{SA}). 
The solid circles on the SC side, on the other hand, give the ${T_c(E_g)}$ 
deduced from the Thouless criterion (\ref{Thcrit}) for BEC. The two critical 
temperatures coincide almost perfectly supporting our claim that even
for finite temperatures the meanfield 
theory captures BEC on the SC side. The small deviations between the two 
curves as we come closer to the SC-SM transition (thick dashed line) 
originate from the discrepancy between the Hulthen and the
screened Coulomb potential when the screening wave number increases.
 
From Fig.~\ref{Fig2} we see that above $T\approx 0.12$ the EI is unstable and an 
ordinary SC-SM transition occurs (thick dashed line), here defined by the vanishing 
of the renormalized exciton binding energy $\bar{B}$. For $T<0.12$, we find a
steeple-like phase boundary, which strongly discriminates between ${E_g}>0$ 
and ${E_g}<0$. For ${E_g}>0$, ${T_c }({E_g}$) first increases 
rather rapidly with increasing ${E_g}$, reaches a maximum, and then decreases 
to zero at ${E_g}=1$,
the critical energy gap, above which the EI cannot exist~\cite{KK64,desC65,KM65,JRK67,HR68}.
For ${E_g}<0$, in contrast, ${T_c(E_g)}$ initially drops relatively fast, 
stays almost constant in a narrow ${E_g}$-range, and then decreases monotonously
with further decreasing ${E_g}$. Notice, in contrast to the qualitative phase boundary 
of Fig.~\ref{Cartoon}, the calculated ${T_c(E_g)}$ reaches the BCS asymptotics 
only at very large band overlaps, far away from the SC-SM transition. On the
scale of the exciton binding energy, we find no simple exponential dependence 
of the transition temperature on the band overlap $|E_g|$. 


The steeple-like shape of the phase boundary reflects the different character of the 
EI when it is approached from the SM and the SC side, respectively. Deep on the SM side
($E_g<0$ and $|E_g|\gg 1$, not shown in Fig.~\ref{Fig2}), 
the EI constitutes a BCS condensate of loosely bound electron-hole pairs whose small 
binding energies determine moreover the low transition temperatures. In contrast, 
on the SC side, the EI is a BEC of tightly bound excitons. The higher transition 
temperatures ${T_c}({E_g})$ on this side are however not determined by the 
larger binding energy but by the temperature for which the ${\bf q}=0$ exciton 
state becomes macroscopically occupied (Thouless criterion (\ref{Thcrit}); solid 
circles). The binding energies per se set only the scale for the SC-SM 
transition~\cite{NSR85}.
\begin{figure}[t]
\includegraphics[width=0.90\linewidth]{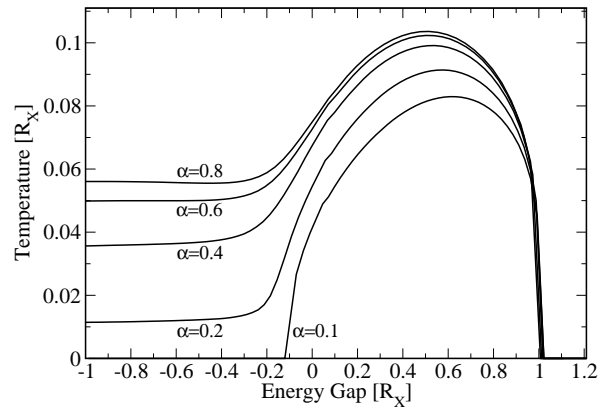}
\caption{\label{Fig3} Phase boundaries for an EI as a function of $\alpha=m_1/m_2$.
For $\alpha\ll 1$, the EI is strongly suppressed on the SM side because
the finite temperature chemical potentials for CB electrons
and VB holes are different for asymmetric band masses leading to breaking of
Cooper-type electron-hole pairs. On the SC side, on the other hand,
transition temperatures are only moderately reduced because of the
$1/M$ dependence typical for BEC.}
\end{figure}

The SC-SM transition is driven by screening and Pauli blocking due to free charge 
carriers which, because of thermal excitation across a small energy gap, are not 
only available on the SM but also on the SC side. Neglecting on the SC side 
thermally excited charge carriers would lead to a much steeper phase boundary 
(dotted line in Fig.~\ref{Fig2}) and to a ${T_c}(0)\approx 0.45$, accidentally 
identical to the guess given for Sr in Ref.~\cite{JRK67}. The relative importance
of phase space filling vs. screening can be estimated by comparing in Fig.~\ref{Fig2}
the SC-SM boundary (thick dashed line) with the hypothetical boundary (thin dashed line)
arising when only screening is taken into account. In that case, the 
SC-SM boundary is given by $B=0$, which is equivalent to the ordinary Mott criterion
$a_X\cdot q_s=1$ (see Eq.~(\ref{Bbare})). Contrasting the thin and thick dashed lines 
demonstrates that Pauli blocking cannot be ignored.

Phase boundaries for asymmetric band masses ($\alpha\neq 1$) are shown in 
Fig.~\ref{Fig3}. The differences of the condensation mechanisms on the SC
and the SM side, respectively, can be most clearly seen in this figure, because 
the transition temperature of a BCS condensate of electron-hole pairs strongly 
depends on $\alpha$ whereas a BEC condensate of excitons does not. In contrast
to equal band masses ($\alpha=1$), where the chemical potentials for the CB electrons 
and VB holes are pinned to $-\bar{E}_{g}/2$, 
as can be seen by inspection from Eqs.~(\ref{neutral}) and~(\ref{Egbar}), asymmetric 
band masses lead 
to chemical potentials which are different for finite temperatures. Band asymmetry 
has thus a pair breaking effect on Cooper-type electron-hole pairs, similar to the 
effect a magnetic field has on Cooper pairs in superconductors.  Accordingly, it 
leads to a strong suppression of ${T_c}({E_g})$ on the SM side. On the 
SC side, on the other hand, the EI is supported by tightly bound excitons and 
the 1/M dependence of the BEC transition temperature leads only to a moderate
$\alpha$ dependence of ${T_c}({E_g})$. Thus, at $\alpha\neq 1$ the
calculated phase boundary deviates substantially from the qualitative guess 
given in Fig.~\ref{Cartoon}.     
\begin{figure}[t]
\includegraphics[width=0.90\linewidth]{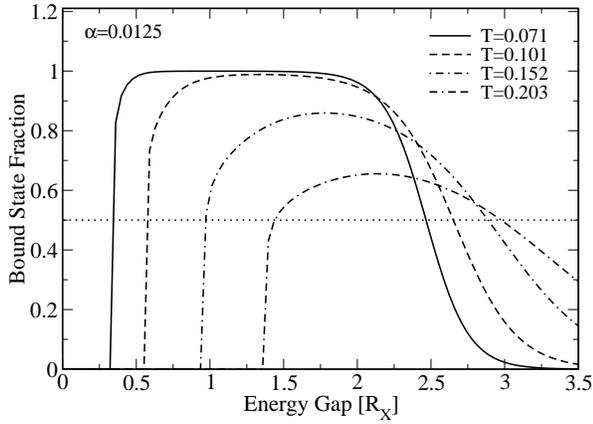}
\caption{\label{Fig4} Bound state fraction of the exciton matter above an
EI with $\alpha=0.0125$ as a function of the energy gap and for
various temperatures. The interceptions with the
dotted line indicate the energy gaps for which the bound state fraction is
for the given temperature 50$\%$.}
\end{figure}

Our results seem to suggest that $\alpha\ll 1$ completely destroys the EI on the
SM side even at ${T=0}$. This is however an artifact of our numerics which reaches
here its limits of accuracy. The EI should be stable at $T=0$ (but not ${T} \neq 0$)
for arbitrary mass ratios $\alpha$. Only anisotropies~\cite{Zittartz67a} and/or
(multi-)valleys in the band dispersions separated by $2{\bf w}\neq{\bf K}$ with
${\bf K}$ a reciprocal lattice vector~\cite{HR68} 
lead at $T=0$ to a suppression of the EI on the SM side for energy overlaps larger 
then a critical value. Note, multi-valleys in a band which are connected by  
a reciprocal lattice vector ${\bf K}$, i.e. identical multi-valleys,  
induce an ``effective'' mass asymmetry~\cite{Zimmermann}
and thus the same pair breaking effect for finite temperatures as the real mass 
asymmetry considered above.

For $\alpha\ll 1$ the EI is almost entirely located on the SC side of
the SC-SM boundary and within our approximations a BEC of excitons. 
Ignoring biexcitons, which, for strong mass asymmetry and attractive 
interactions, could interfere on the SC side with the EI~\cite{HR68},
two temperatures are required to characterize the EI and its exciton 
environment (halo of an EI):
${T_M}({E_g})$ where 
exciton formation sets in and ${T_c}({E_g})$ where
condensation (phase coherence) is finally reached. Above ${T_c}({E_g})$
but below ${T_M}({E_g})$, a mixture of excitons, free electrons, and
free holes exists, the composition of which depends on $T$ and ${E_g}$,
as can be seen in Fig.~\ref{Fig4}, where we plot bound state fractions
\begin{eqnarray}
\gamma=\frac{n_1^{b}}{n_1^{f}+n_1^{b}},
\label{IonDegree}
\end{eqnarray}
above ${T_c}({E_g})$ for an EI with $\alpha=0.0125$. Here,
$n_1^{f}$ denotes the part of the total density corresponding to free 
electrons and $n_1^{b}$ the part bound in electron-hole pairs
(excitons) as given by Eqs.~(\ref{n1fApp}) and (\ref{n1cApp}), respectively. 
\begin{figure}[t]
\includegraphics[width=0.90\linewidth]{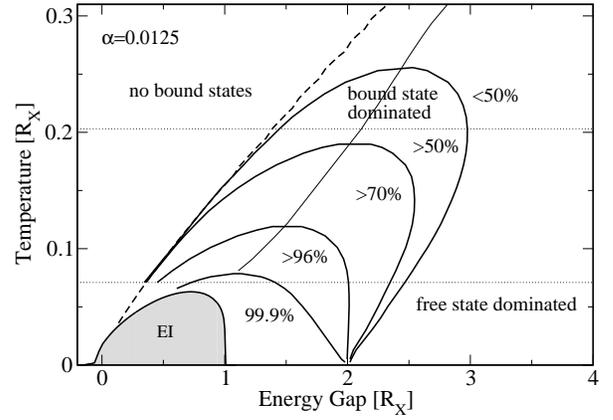}
\caption{\label{Fig5} Phase boundary of an EI with $\alpha=0.0125$
and contours where the bound state fraction of its halo is
99.9$\%$, 96$\%$, 70$\%$, and 50$\%$, respectively. The temperature-dependent
position of the maximum of the bound state fraction is shown by the thin solid
line. Thin dashed lines indicate the temperatures for which in
Fig.~\ref{DenPlot} the densities are plotted.}
\end{figure}

The bound state fraction indicates on which side of the chemical equilibrium
$e+h\rightleftharpoons X$ the system is according to the mass-action law~\cite{ZS85}.
Deep on the SC side, the chemical 
equilibrium is practically for all temperatures on the side of unbound
electrons and holes and the bound state fraction is zero. Decreasing the
energy gap leads to a shift of the chemical equilibrium to the exciton
side signalled by an increasing bound state fraction which assumes a temperature
dependent maximum (which can be very close to unity) before it abruptly decreases again 
to zero at and beyond the SC-SM transition. The depth and width of the 
maximum strongly depends on temperature. At low temperatures and large
energy gaps the bound state fraction acquires a step-like shape 
(already visible by the solid line in Fig.~\ref{Fig4}). In that parameter 
range, the exponential tails of the Fermi and Bose functions 
contribute to Eqs.~(\ref{n1fApp}) and (\ref{n1cApp}). Moreover, 
$\bar{B}\rightarrow B\rightarrow R_X$, $\bar{E}_g\rightarrow E_g$, 
and $\mu_1\approx\mu_2\approx -E_g/2$, leading to
\begin{eqnarray}
\gamma\rightarrow 1-\frac{1}{1+(\frac{1+\alpha}{\alpha})^{\frac{3}{2}}\exp{[\beta(R_X-E_g/2)}]}. 
\label{asymp}
\end{eqnarray}
Thus, as a consequence of the mass-action-law, the bound state fraction 
$\gamma$ has a step at $E_g=2\cdot R_X$ for $\beta E_g, \beta(E_g-R_X)\ll 1$.

The phase boundary for the EI with $\alpha=0.0125$ together with  
contours of the bound state fraction of its halo are displayed in
Fig.~\ref{Fig5}. Almost the whole EI sits on the SC side, where it is surrounded 
by a large exciton rich region with bound state fractions significantly 
above 50$\%$. Remarkably, the bound state fraction approaches almost  
100$\%$ {\it before} phase coherence is established. The exciton density in that
region is not yet large enough for condensation to occur. Notice also, 
as a consequence of Eq.~(\ref{asymp}), all contours in Fig.~\ref{Fig5} start
at ${E_g}=2$ (exciton units).
\begin{figure}[t]
\includegraphics[width=0.90\linewidth]{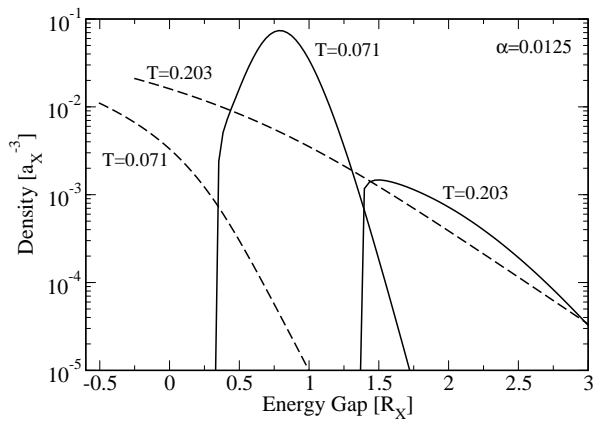}
\caption{\label{DenPlot} Bound (solid line) and free parts (dashed line) of the
CB electron density as a function of the energy gap and for two temperatures
above the EI (dotted lines in Fig.~\ref{Fig5}). Notice the very sharp decrease
of the bound part of the density when the SC-SM transition is approached.}
\end{figure}

Figure~\ref{Fig5} shows the exciton dominated part of the halo 
($\gamma\ge 0.5$). In that region, we expect electrical 
resistivity anomalies similar to the ones observed in ${\rm TmSe_{0.45}Te_{0.55}}$~\cite{NW90,BSW91}. 
Above ${T_c}({E_g})$ but below ${T_M}({E_g})$, for instance at $T=0.071$, diminishing ${E_g}$
(increasing pressure) pushes the system from the free state dominated to the bound 
state dominated regime. In the former, the resistivity is expected to decrease with
decreasing ${E_g}$, because decreasing ${E_g}$ leads to an increase 
of free electrons and holes. The resistivity is here determined by the scattering 
of free charge carriers on imperfections and phonons. In the latter, however, decreasing 
${E_g}$ leads not only to an increase of the free charge carriers but also to 
a rather strong increase of bound states, as can be seen in Fig.~\ref{DenPlot}. Depending on 
temperature and $E_g$, the bound part of the density can be orders of magnitude higher 
then the density of the free charge carriers responsible for charge transport. As a result, 
an additional scattering channel is now available, free--bound state scattering, 
and strongly increases the resistivity. After 
a temperature dependent critical ${E_g}$, the bound part 
of the density decreases with decreasing energy gap. Hence, free--bound state scattering
diminishes and the resistivity decreases again until it changes abruptly to a lower
value at the SC-SM transition. 

The results we presented in this section are for a generic Wannier-type EI.
Contact with particular materials can be made by specifying the mass ratio 
$\alpha$, the exciton Rydberg $R_X$, and the exciton Bohr radius $a_X$ (or, 
alternatively, one of the two effective band masses $m_i$).

\section{\label{Comparision}IV. Comparison with Experiment}

We now discuss from the vantage point of our theory the experimental data for the
pressure-induced SC-SM transition in
${\rm TmSe_{0.45}Te_{0.55}}$~\cite{NW90,BSW91,W01,WBM04}. Because of the fcc
crystal structure~\cite{NW90}, the VB has a single maximum at the $\Gamma$ point while
the CB has minima at the three $X$ points of the Brillouin zone. We can formally account
for the three identical valleys of the CB as well as for the spin degeneracy by
introducing on the rhs of Eqs. (\ref{n1}) and (\ref{n2}) multiplicity factors $g_1=6$
and $g_2=2$, respectively. The remaining equations are unchanged.

The mass ratio estimates for ${\rm TmSe_{0.45}Te_{0.55}}$ range from $\alpha=0.01$
to $\alpha=0.02$~\cite{WBM04} and optical measurements reveal an energy gap
of 135~meV and an exciton binding energy of 75~meV~\cite{NW90}. The binding energy
is thus rather large, which, in addition to the high exciton densities (see below), 
indicates that the effective-mass, Wannier-type model may not be quite adequate.
Hybridization effects characteristic for a mixed-valence material such as 
${\rm TmSe_{0.45}Te_{0.55}}$ may also not be completely captured by a two-band
model. For our model to mimic the relevant parts of the band structure
of ${\rm TmSe_{0.45}Te_{0.55}}$ we have to assume that during the 
excitonic instability, hybridization does not change much and the composition 
of the two bands is more or less fixed. 

Although the multi-valley CB should favor a metallic electron-hole 
liquid~\cite{R77}, the experimentally observed increase of the resistivity 
cannot be explained by an emerging metallic phase. We consider therefore 
the static screening model, which a priori favors excitonic phases, 
as an effective model for ${\rm TmSe_{0.45}Te_{0.55}}$. The investigation
of the competition between the two phases should be based on an improved  
description of screening (see the discussion following 
Eqs. (\ref{SN})-(\ref{Qgap})), taking however additional scattering processes
into account, which most probably stabilize the EI in ${\rm TmSe_{0.45}Te_{0.55}}$ 
against an electron-hole liquid. Exciton-phonon scattering, for instance, could
be such a process because the narrow VB in ${\rm TmSe_{0.45}Te_{0.55}}$
makes the exciton very susceptible to phonon dressing. Indeed, phonon 
signatures have been experimentally seen~\cite{WBM04}, but it is not clear 
whether they are the driving force for the excitonic instability or only 
triggered by it. In this respect it should be also mentioned that a static 
distortion associated with a density wave (which would be expected 
from the indirect gap) has not been found experimentally.

With these restrictions in mind, we present in Fig.~\ref{Fig6} the phase boundary
for $\alpha=0.015$, calculated with the multiplicity factors given above and 
scaled with the exciton Rydberg applicable to ${\rm TmSe_{0.45}Te_{0.55}}$, and 
compare it with the phase boundary for the ``excitonic insulator'' given by Wachter 
and coworkers in Ref.~\cite{BSW91}. Recall the phase boundary was constructed
from the anomalous behavior of the electrical resistivity which, in a
particular pressure and temperature range, first increases, peaks at a
certain pressure, and then decreases with pressure until it terminates in a
discontinuous jump. The on-set of the increase and the jump have been used to
determine the entrance points for the ``excitonic insulator'' from the SC (open circles)
and the SM from the ``excitonic insulator'' (open triangles), respectively. The ${E_g-T}$
values where the resistivity is maximal are given by the open squares. Fig.~\ref{Fig6} also
shows the data point (full diamond on the dotted line) where the linear increase of the
thermal diffusivity has been observed. Wachter and coworkers claim that at this point
${\rm TmSe_{0.45}Te_{0.55}}$ is in a superfluid exciton state~\cite{WBM04}.
\begin{figure}[t]
\includegraphics[width=0.90\linewidth]{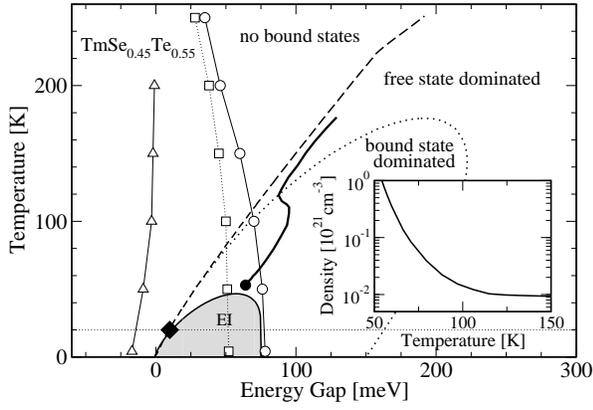}
\caption{\label{Fig6} Comparison of the experimental data 
for ${\rm TmSe_{0.45}Te_{0.55}}$~\cite{BSW91,WBM04} with 
the phase boundary and the 50$\%$ contour calculated for an effective 
model applicable to this substance (see text for a description).
Open circles, squares, and triangles denote the onset of the
resistivity anomaly, the maximum of the resistivity, and the SC-SM
transition, respectively~\cite{BSW91}. The full diamond on the
dotted line indicates the point where the linear increase of the
diffusivity has been found in Ref.~\cite{WBM04}. The inset shows
the bound part of the density, $n_1^b$, along the thick solid line, which
denotes the $(E_g,T)$ points for which $n_1^b$ is maximal. At the
full circle $n_1^b\approx 1.3\times 10^{21}~{\rm cm}^{-3}$.}
\end{figure}

Apparently, Wachter and coworkers~\cite{WBM04} distinguish between an ``excitonic
insulator'' at high temperatures and a superfluid exciton state at low temperatures
which, however, is misleading, because the term ``excitonic insulator'' has been originally
introduced to describe, on the SM side, a superfluid electron-hole condensate, similar to a 
superfluid condensate of Cooper pairs in a superconductor, and, on the SC side, an
exciton condensate, similar to superfluid He~4~\cite{desC65,KK64,KM65,JRK67,HR68}.
Although the precise meaning of the terms ``excitonic insulator'' and ``exciton condensation''
changed in the course or their work and, in fact, was never fully in accordance with 
the original definitions, in more recent publications~\cite{W01,WBM04} they  
denote by the former an exciton liquid and by the latter a gas/liquid transition. 
The transition to the superfluid state they consider to be similar 
to the one in liquid He~4~\cite{WBM04}.

Based on the results presented in the previous section, we suspect that the resistivity 
anomaly in ${\rm TmSe_{0.45}Te_{0.55}}$ is due to free--bound state scattering in the 
exciton rich part of the halo of
an EI whereas the linear increase of the thermal diffusivity could signal an EI, 
understood as a macroscopic, phase coherent condensate of excitons. If our hypotheses 
are correct, we should be able to (i) relate the 
phase boundary of the ``excitonic insulator'' to the above 50$\%$ contours of the 
bound state fractions and (ii) find the experimental signature for superfluidity at
a temperature where we would expect the EI. 

The experimentally determined exciton Rydberg for ${\rm TmSe_{0.45}Te_{0.55}}$ 
puts the exciton rich part of the halo in the temperature range $T\sim$~50 -- 180~K, 
close to the temperature range where the resistivity anomaly and thus the 
``excitonic insulator'' has been 
observed (open triangles, squares, and circles in Fig.~\ref{Fig6}). The 
bound part of the density in this range has also the correct order of magnitude. At 
$T=53~{\rm K}$ and ${E_g}=64~{\rm meV}$ (full circle in Fig.~\ref{Fig6}), for instance,
we find, using $m_1=2.1m_e$ for the CB mass ($m_e$ bare electron mass), 
$n^b_1\approx 1.3\times 10^{21}~{\rm cm}^{-3}$ 
which compares favorably with the experimentally estimated exciton density of 
$3.9\times 10^{21}~{\rm cm}^{-3}$~\cite{W01}. The bound part of the density reaches 
its maximum along the thick solid line in Fig.~\ref{Fig6}. As can be seen in the 
inset, below 100~K, i.e. within the bound state dominated region, 
$n^b_1$ increases sharply. An improved theory, taking interactions 
between electron-hole pairs into account, could, when the repulsive part of the 
interaction dominates, display a gas/liquid transition in 
that temperature range. Hence, an exciton fluid, as proposed in Ref.~\cite{WBM04}, 
is conceivable below 100~K. 

Below 50~K we find the EI, that is, the macroscopic, phase coherent condensate, 
very close to the 20~K (dotted line) where the linear increase of the 
diffusivity was observed at 13~kbar. From the associated isobar, we 
estimate the energy gap for that data point to be $\approx$~20~meV (full 
diamond). Wachter and coworkers' claim that this increase is due to the second 
sound of a superfluid is supported by our theory because the 
EI is entirely on the SC side. Hence, it constitutes a BEC of bosonic excitons, and 
not a BCS condensate of loosely bound electron-hole pairs, which may~\cite{KM66b} or 
may not display superfluidity~\cite{Zittartz68b}. Without detailed knowledge 
of the interaction between electron-hole pairs, we cannot calculate superfluid 
properties. Nevertheless, the fact that the EI is on the BEC side makes it plausible
that an improved theory would yield superfluidity, provided the 
interaction between excitons is dominantly repulsive.

The temperature dependence of the bound state fractions 
is consistent with values obtained form ab-initio numerical simulations of
mass asymmetric electron-hole plasmas~\cite{FFBFL05}. The approximate
T-matrix (\ref{vertex}) on which the calculation of the bound state 
fractions is based captures therefore the essential physics.  
The interpretation of the experimental data is however hampered by two 
main difficulties. 

First, the effective-mass model~(\ref{model}) and the resulting Wannier-type 
EI are, strictly speaking, not applicable to ${\rm TmSe_{0.45}Te_{0.55}}$.
The exciton binding energy is too large compared to the energy gap. As a 
result, the bound state dominated region in Fig.~\ref{Fig6} completely 
exhausts the maximal energy gap of 135~meV available at ambient pressure. 
Hence, the free state dominated region, where we expect a decrease of the 
resistivity with decreasing energy gap, occurs at energy gaps which cannot
be realized.

Second, the bare energy gaps $E_g$, which we employ as a control
parameter, are not necessarily the energy gaps given by Wachter and coworkers. 
Using the linear relation ${E_g(p,T=300~{\rm K})}={E_g(p=p_{a},T=300~{\rm K})}
+{(dE_g/dp)(T=300~{\rm K})(p-p_{a})}$, with ${p_{a}}$ the 
ambient pressure, they convert (pressure, temperature) points, which they 
control experimentally, to (energy gap, temperature) points assuming
the energy gap to be temperature independent and the closing rate to be 
temperature and pressure independent~\cite{NW90,BSW91}.
At finite temperatures, however, bands are partially filled. The band 
gap renormalization to which the partial filling gives rise to leads not
only to a strong temperature dependence of the energy gap but most probably
also to a complicated pressure and temperature dependence of the 
closing rate. These effects were not incorporated in the assignment 
(pressure, temperature) $\leftrightarrow$ (energy gap, temperature) and 
perhaps the reason for the slight `left-turn' of the experimental curves. 

\section{\label{Conclusion}V. Conclusions}
We adopted a BEC-BCS crossover scenario to calculate the phase boundary 
of a pressure-induced EI using an isotropic, effective-mass  
model to describe the two bands involved in the excitonic instability
and the dominant-term-approximation for the (statically screened) Coulomb 
interaction.

Assuming pressure to affect only the energy gap, we derived, within 
a meanfield approximation, a linearized gap equation for the
anomalous selfenergy (order parameter) and determined, for a given  
energy gap ${E_g}$, the transition temperature ${T_c(E_g)}$ as the highest 
temperature for which a non-trivial solution exists. The kernel of the gap
equation depends on the chemical potentials for CB electrons and VB holes
and on the screening wave number, all three have been determined
self-consistently as a function of $T$ and ${E_g}$. 

We pointed out that because pressure directly controls the chemical 
potentials, which moreover are constrained by charge neutrality, the meanfield 
theory is even for finite temperatures on par with T-matrix based crossover 
theories which include pair states in the equation for the chemical 
potential. We demonstrated by physical reasoning and calculation 
that in our case, pair states do not affect the chemical potentials as long
as one stays within the framework of noninteracting pairs. As a result, the 
transition temperatures obtained from our meanfield theory coincide on the SC 
side with the transition temperatures for a BEC of non-interacting bosonic excitons. 
As a consistency check, we also determined ${T_c(E_g)}$ on the SC side from 
the normal phase electron-hole T-matrix using the Thouless criterion.

Our main results are as follows: For equal band masses we found a very asymmetric 
phase boundary 
${T_c(E_g)}$, with a steeple on the SC side and a long tail on the SM side,
directly reflecting the different character of the EI on the SC and SM
side, respectively. Whereas on the SM side the EI comprises a BCS condensate 
of loosely bound Cooper pairs, the EI is a BEC of tightly bound excitons
on the SC side. On the SM side, no simple exponential dependence
of the transition temperature on the band overlap $|E_g|$ exists, as perhaps
expected from the formal analogy to the BCS theory of superconductivity. We 
found mass asymmetry to strongly suppress  
pairing on the SM side because the different temperature dependencies of the 
chemical potentials for the CB electrons and VB holes have a pair-breaking 
effect on Cooper-type electron-hole pairs. The
transition temperatures on the SC side, on the other hand, are only moderately 
modified, because the center-of-mass motion of excitons, which determines the 
transition temperatures on that side, depends only algebraically on the total 
mass (${T_c}\sim M^{-1}$). For very  
asymmetric band masses the EI is therefore rather fragile on the SM side and
for finite temperatures almost entirely located on the SC side of the SC-SM transition.
On the SC side, we also investigated the exciton density (more precisely, the 
part of the total density bound in excitons) in the region above 
$T_c$ which we called halo. Because of the mass-action-law, the exciton 
density strongly depends on temperature and energy gap. As a consequence, we 
expect pronounced resistivity anomalies in the halo due to free--bound state 
scattering. 

Within the limitations of our model, we attempted an analysis of the 
experimental data for the pressure-induced SC-SM transition in 
${\rm TmSe_{0.45}Te_{0.55}}$~\cite{NW90,BSW91,W01,WBM04}. From our 
theoretical point of view, the phase boundary, which was 
constructed from electrical resistivity data, traces most probably the exciton 
rich surroundings of an EI, where excitons dominate the total density, provide
abundant scattering centers for the charge carriers, and induce the observed
resistivity anomaly. Since we neglect interactions between excitons, we cannot 
decide whether a gas/liquid transition is also involved in the resistivity 
anomaly~\cite{WBM04}. The linear increase of the thermal diffusivity which 
was taken as a signal for superfluidity occurs at a temperature where 
we would expect the EI. Most importantly, however, the EI and the data point 
are on the SC side of the SC-SM transition, where the EI, that is, the 
macroscopic, phase coherent exciton state, 
consists of tightly bound bosonic excitons, which should support
superfluidity when the repulsive core of the exciton-exciton interaction 
prevails over the attractive tail. Thus, the long journey of Wachter and 
coworkers~\cite{NW90,BSW91,W01,WBM04} may indeed 
have culminated~\cite{WBM04} in the first observation of an exciton 
condensate.

We could not unambiguously proof the existence of a superfluid exciton condensate
in ${\rm TmSe_{0.45}Te_{0.55}}$, most notably, because we did not 
allow the EI to compete with a metallic electron-hole liquid and because we 
did not account for the possibility of density waves. Although neither is
supported by experimental data, both are in principle possible in a 
multi-valley band structure. Thus, to further substantiate the experimentalists
claim from the theoretical side, a unified treatment of excitonic and electron-hole
liquid phases, accounting also for the possibility of density waves, is clearly 
necessary, ideally for a model which avoids the effective-mass approximation, 
includes exciton-phonon scattering (which could stabilize excitonic phases
against an electron-hole liquid),  
and an approximation scheme which takes the interaction between electron-hole
pairs (or excitons) into account. The latter is important for the calculation 
of transport coefficients, which could be directly compared with 
experimental data, as  well as for clarifying the role of biexciton
formation, which, for large mass asymmetry, competes with the formation of the EI.
Besides diagrammatic techniques~\cite{Roepke94} and systematic 
mode-coupling approaches~\cite{CG88,CG93}, numerical simulation techniques~\cite{SC99,FFBFL05}
could proof very useful in this respect. In any case, the comparison between 
theory and experiment depends on the assignment (pressure, temperature)
$\leftrightarrow$ (energy gap, temperature), which is rather subtle. To 
study the SC-SM transition along `iso-$E_g$' instead of isobars would 
eliminate the uncertainties associated with this critical assignment and 
thus open the door for a more rigorous analysis of the excitonic phases in 
${\rm TmSe_{0.45}Te_{0.55}}$. 

\appendix*
\section{Appendix A: CB electron and VB hole densities in T-matrix approximation}

In this appendix we determine the CB electron and VB hole densities as a function of
the bare energy gap and the temperature. Thereby, we verify that even in the T-matrix 
approximation the chemical potentials for CB electrons and VB holes are, in leading
order, not affected by exciton states.

Our starting point are the well-known formulae
\begin{eqnarray}
n_1&=&\int \frac{d{\bf k}}{(2\pi)^3} \frac{d\omega}{2\pi} A_{11}({\bf k},\omega)n_{\rm F}(\omega),
\label{densityelectron}\\\nonumber\\
\bar{n}_2&=&\int \frac{d{\bf k}}{(2\pi)^3} \frac{d\omega}{2\pi} A_{22}({\bf k},-\omega)n_{\rm F}(\omega) 
\label{densityhole}
\end{eqnarray}
for the CB electron and VB hole density, respectively. The spectral functions $A_{ii}$ are 
calculated with the selfenergies shown in Figs.~(\ref{Fig1}a) and (\ref{Fig1}b). Expanding 
the spectral functions with respect to
$\Gamma_{ii}=-2{\rm Im}\Sigma_{ii}$ leads to~\cite{ZS85}
\begin{eqnarray}
A_{ii}({\bf k},\omega)&=&2\pi Z_{ii}({\bf k})\delta(\omega-\bar{e}_i({\bf k}))
\nonumber\\
&+&\Gamma_{ii}({\bf k},\omega)\frac{\partial}{\partial \bar{e}_i({\bf k})}
\frac{\rm P}{\omega-\bar{e}_i({\bf k})}
\label{A11}
\end{eqnarray}
with the quasi-particle weight
\begin{eqnarray}
Z_{ii}({\bf k})&=&1-\frac{\partial}{\partial \bar{e}_i({\bf k})}
{\rm P}\int \frac{d\omega}{2\pi}\frac{\Gamma_{ii}({\bf k},\omega)}{\omega-\bar{e}_i({\bf k})}
\label{Z11}
\end{eqnarray}
and the renormalized single particle dispersion
\begin{eqnarray}
\bar{e}_i({\bf k})={e}_i({\bf k})+\Sigma_{ii}^X({\bf k})+
{\rm Re}\Sigma_{ii}^L({\bf k},\bar{e}_i({\bf k})+i\eta).
\label{eladder}
\end{eqnarray}

Let us first consider the CB electron density. Assuming the internal propagators in the selfenergy 
to describe free quasi-particles with
a renormalized dispersion $\bar{e}_i$ but a quasi-particle weight unity, the selfenergy arising
from the T-matrix reads
\begin{widetext}
\begin{eqnarray}
\Sigma_{11}^L({\bf k},\omega_n)&=&\int\frac{d{\bf q}}{(2\pi)^3}\frac{1}{\beta}\sum_{i\Omega_n}
\Lambda_{12}({\bf k},{\bf k};{\bf q},i\Omega_n){\cal G}_{22}({\bf k}-{\bf q},i\omega_n-i\Omega_n)
\nonumber\\
&=&\int\frac{d{\bf q}}{(2\pi)^3}\big[g(|{\bf k}-\frac{m_1}{M}{\bf q}|)\big]^2 Z_X
\frac{n_{\rm B}(E_X({\bf q}))+n_{\rm F}(\varepsilon_2({\bf k}-{\bf q})-\mu_2)}
{E_X({\bf q})-\varepsilon_2({\bf k}-{\bf q})+\mu_2-i\omega_n},
\label{SigmaL}
\end{eqnarray}
where we introduced the Bose function $n_{\rm B}$ and the exciton dispersion 
$E_X({\bf q})=\bar{E}_{g}-\bar{B}+q^2/2M$.
To obtain Eq.~(\ref{SigmaL}), we approximated the renormalized dispersions $\bar{e}_i$,
that is, the solutions of Eq.~(\ref{eladder}), by  Eq.~(\ref{rendis}). For the electron-hole
T-matrix we used Eq.~(\ref{vertex}) with Eq.~(\ref{Xprop1}) for the renormalized exciton 
propagator. Thus, after analytical continuation to real frequencies, we obtain
\begin{eqnarray}
\Gamma_{11}({\bf k},\omega)=\int\frac{d{\bf q}}{(2\pi)^2}
\big[g(|{\bf k}-\frac{m_1}{M}{\bf q}|)\big]^2 Z_X \big[n_{\rm B}(E_X({\bf q}))+
n_{\rm F}(\varepsilon_2({\bf k}-{\bf q})-\mu_2)\big]
\delta(E_X({\bf q})-\varepsilon_2({\bf k}-{\bf q})+\mu_2-\omega)
\label{Gamma11}
\end{eqnarray}
\end{widetext}
for the imaginary part of the selfenergy on the real frequency axis and, using 
a Kramers-Kronig relation,  
\begin{eqnarray}
{\rm Re}\Sigma^L_{11}({\bf k},\omega)={\rm P}\int\frac{d\omega'}{2\pi}
\frac{\Gamma_{11}({\bf k},\omega')}{\omega-\omega'},
\label{ReS11}
\end{eqnarray}
for the real part of the selfenergy on the real frequency axis.

If we now insert Eqs.~(\ref{Gamma11}) and (\ref{ReS11}) into Eq.~(\ref{A11}),
recall once again Eq.~(\ref{rendis}), and perform the ${\bf k}$-integration in 
Eq.~(\ref{densityelectron}), we find for the CB electron density
\begin{eqnarray}
n_1=n_1^f+n_1^b,
\label{n1App}
\end{eqnarray}
with a term arising from free electrons (in the sense defined above)
\begin{eqnarray}
n_1^f=\int\frac{d{\bf k}}{(2\pi)^3}n_{\rm F}(\varepsilon_1({\bf k})-\mu_1)
\label{n1fApp}
\end{eqnarray}
and a term 
\begin{eqnarray}
n_1^b=\int\frac{d{\bf p}d{\bf q}}{(2\pi)^6}
\bigg[\frac{B+\frac{p^2}{2m}}{\bar{B}+\frac{p^2}{2m}}\bigg]^2
\chi^2(p) Z_X F_1({\bf p},{\bf q})
\label{n1cApp}
\end{eqnarray}
due to the coupling of CB electrons and VB holes. The statistical factor
is given by
\begin{eqnarray}
F_1({\bf p},{\bf q})&=&
n_{\rm B}(E_X({\bf q}))\big[1-n_{\rm F}(\varepsilon_1({\bf p}+\frac{m_1}{M}{\bf q})-\mu_1)
\nonumber\\
&-&n_{\rm F}(\varepsilon_2({\bf p}-\frac{m_2}{M}{\bf q})-\mu_2)\big]\nonumber\\
&-&n_{\rm F}(\varepsilon_1({\bf p}+\frac{m_1}{M}{\bf q})-\mu_1)
\nonumber\\
&\times&n_{\rm F}(\varepsilon_2({\bf p}-\frac{m_2}{M}{\bf q})-\mu_2).
\end{eqnarray}
To derive Eq.~(\ref{n1cApp}) we transformed the ${\bf k}$-integration in 
Eq.~(\ref{densityelectron}) to an integration over the relative momentum 
${\bf p}={\bf k}-(m_1/M){\bf q}$ when $\Gamma_{11}$ was part of the integrand and used 
definition (\ref{Wannier}) of the form factor.

In the limit $\beta E_g\ll 1$, Eq.~(\ref{n1cApp}) reduces to 
\begin{eqnarray}
n_1^b=\int\frac{d{\bf q}}{(2\pi)^3}n_{\rm B}(E_X({\bf q}))
\end{eqnarray} 
because $F_1({\bf p},{\bf q}) \rightarrow n_{\rm B}(E_X({\bf q}))$,
$Z_X\rightarrow 1$, and the integration 
over ${\bf p}$ yields unity. The latter, a consequence of the optical theorem
which the T-matrix has to obey, can be only ensured within a skeleton expansion. 

The VB hole density can be calculated in a similar way. The result is 
\begin{eqnarray}
\bar{n}_2=\bar{n}_2^f+\bar{n}_2^b
\label{n2App}
\end{eqnarray}
with
\begin{eqnarray}
\bar{n}_2^f=\int\frac{d{\bf k}}{(2\pi)^3}n_{\rm F}(\varepsilon_2({\bf k})-\mu_2)
\end{eqnarray}
and a bound part $\bar{n}_2^b$ which is identical to 
$n_1^b$ given in Eq.~(\ref{n1cApp}). This is not unexpected, because 
for each CB electron participating in a bound state, a VB hole has to do so too. 

As a consequence, the charge neutrality condition, $n_1=\bar{n}_2$, which fixes the 
chemical potentials for CB electrons and VB holes, reduces to $n^f_1=\bar{n}^f_2$. 
Neglecting the modifications of the single particle dispersions due to excitons,
that is, ignoring interactions between excitons, which are higher order effects, 
the charge neutrality condition in the T-matrix approximation reduces to the meanfield 
charge neutrality condition. 

With the chemical potentials determined from Eqs.~(\ref{neutral}) and (\ref{Egren}), 
Eqs.~(\ref{n1fApp}) and (\ref{n1cApp}) can be used to calculate, above the 
critical temperature ${T_c(E_g)}$, the free and bound parts of the CB electron 
density and thus the bound state fraction (\ref{IonDegree}).

\section{Appendix B: Separable approximation for the normal phase electron-hole T-matrix}
Here we derive a separable approximation for the normal phase electron-hole 
T-matrix $\Lambda_{12}({\bf k,k'};{\bf q},i\Omega_n)$. In the vicinity 
of the exciton state the T-matrix factorizes with respect to ${\bf k}$ and 
${\bf k'}$. Thus, the separable approximation is a good approximation for energies 
not too far from the exciton energy. Since this energy range determines the
physics of an EI, the separable approximation should be sufficient for our 
purpose.

The approximation is based on an expansion of the normal phase electron-hole T-matrix 
in terms of the eigenfunctions of an auxiliary Lippmann-Schwinger equation which takes 
into account screening due to spectator particles not participating in the bound state 
but ignores the spectators' Pauli blocking. The phase space filling responsible for 
Pauli blocking is incorporated in a second step through a selfenergy for the 
exciton propagator.  
To avoid the numerical solution of the auxiliary Lippmann-Schwinger equation we 
approximate, as far as the calculation of the T-matrix is concerned,
the screened Coulomb potential by the Hulthen potential. Eigenfunctions and binding
energies are then analytically available. 

\begin{widetext}
The starting point is Eq.~(\ref{LadderEq}). In a first step, we break up the screened 
ladder equation into two equations. In an obvious notation
\begin{eqnarray}
\Lambda_{12}&=&\Lambda_{12}^{(0)}+\Lambda_{12}^{(0)}[Q-Q^{(0)}]\Lambda_{12},\label{vertex1}\\
\Lambda_{12}^{(0)}&=&V_s+V_s Q^{(0)}\Lambda_{12}^{(0)}\label{vertex0}
\end{eqnarray}
with 
\begin{eqnarray}
Q^{(0)}({\bf k};{\bf q},i\Omega_n)=
\frac{1}
{\bar{E}_{g}+\varepsilon_1({\bf k})+\varepsilon_2({\bf k}-{\bf q})-i\Omega_n}.
\end{eqnarray}
Note, this is an exact rewriting of Eq.~(\ref{LadderEq}), with the advantage that screening and 
Pauli blocking are now treated separately. Equation~(\ref{vertex0}) introduces 
an auxiliary vertex which considers screening but not Pauli blocking. The full vertex is 
then obtained from Eq.~(\ref{vertex1}). 

Introducing relative momenta ${\bf p}$ 
according to ${\bf k}={\bf p}+(m_1/M){\bf q}$,
Eq.~(\ref{vertex0}) becomes 
\begin{eqnarray}
\Lambda_{12}^{(0)}({\bf p}+\frac{m_1}{M}{\bf q},{\bf p'}+\frac{m_1}{M}{\bf q};{\bf q},i\Omega_n)
\!\!&=&\!\!V_s({\bf p}-{\bf p'})\nonumber\\
\!\!&+&\!\!\int \frac{d{\bf p''}}{(2\pi)^3} 
\frac{V_s({\bf p}-{\bf p''})}
{\bar{E}_{g}+\frac{q^2}{2M}+\frac{p''^2}{2m}-i\Omega_n}
\Lambda_{12}^{(0)}({\bf p''}+\frac{m_1}{M}{\bf q},{\bf p'}+\frac{m_1}{M}{\bf q};{\bf q},i\Omega_n),
\end{eqnarray}
which, with the identification
\begin{eqnarray}
\Lambda_{12}^{(0)}({\bf p}+\frac{m_1}{M}{\bf q},{\bf p'}+\frac{m_1}{M}{\bf q};{\bf q},i\Omega_n)
=T({\bf p},{\bf p'};i\Omega_n-\bar{E}_{g}-\frac{q^2}{2M}),
\label{identification}
\end{eqnarray}
reduces to 
\begin{eqnarray}
T({\bf p},{\bf p'};i\Omega_n)
=V_s({\bf p}-{\bf p'})
+\int \frac{d{\bf p''}}{(2\pi)^3}
\frac{V_s({\bf p}-{\bf p''})}
{\frac{p''^2}{2m}-i\Omega_n}
T({\bf p''},{\bf p'};i\Omega_n).
\label{Tmatrix}
\end{eqnarray}
Equation~(\ref{Tmatrix}) is the T-matrix equation in the center-of-mass frame of an 
electron-hole pair interacting via a screened Coulomb potential. The medium occurs
here only through the screening wave number $q_s$. 

To solve Eq.~(\ref{Tmatrix}) we employ the unitary pole expansion for $T({\bf p},{\bf p'};i\Omega_n)$,
using eigenfunctions of the corresponding Lippmann-Schwinger equation at fixed energy as a 
basis~\cite{Harms70}. In particular, we take the eigenfunctions at energy 
$i\Omega_n\rightarrow -B$, with $-B$ the energy of the lowest bound state. 
Truncating the expansion after the
first term yields (for an isotropic system)
\begin{eqnarray}
T(p,p';i\Omega_n)=g(p)\cdot d_X(i\Omega_n)\cdot g(p')
\label{UPA}
\end{eqnarray}
with 
\begin{eqnarray}
d_X(z)=\frac{-1}{z+B},
\end{eqnarray}
which can be interpreted as a screened exciton propagator, and form factors defined in 
Eq.~(\ref{Wannier}). For the Hulthen potential the screened exciton wavefunction $\chi(p)$,
and thus the from factor $g(p)$, as well as the screened binding energy $B$ can be obtained 
analytically and are given by Eqs. (\ref{chi}) and (\ref{Bbare}), respectively. 

Combining now Eq.~(\ref{UPA}) with Eq.~(\ref{identification}) leads to
\begin{eqnarray}
\Lambda_{12}^{(0)}({\bf p}+\frac{m_1}{M}{\bf q},{\bf p'}+\frac{m_1}{M}{\bf q};{\bf q},i\Omega_n)
=g(p)\cdot d_X(i\Omega_n-\bar{E}_{g}-\frac{q^2}{2M}) \cdot g(p')
\label{auxiliary}
\end{eqnarray}
for the auxiliary vertex. 
To obtain the full vertex, we write $\Lambda_{12}$ in a similar form, 
\begin{eqnarray}
\Lambda_{12}({\bf p}+\frac{m_1}{M}{\bf q},{\bf p'}+\frac{m_1}{M}{\bf q};{\bf q},i\Omega_n)
=g(p)\cdot D_X({\bf q},i\Omega_n-\bar{E}_{g}-\frac{q^2}{2M})\cdot g(p'),
\label{full}
\end{eqnarray}
but with a renormalized exciton propagator $D_X({\bf q},i\Omega_n)$ instead of the screened 
exciton propagator 
$d_X(i\Omega_n)$. Inserting Eqs.~(\ref{auxiliary}) and~(\ref{full}) into Eq.~(\ref{vertex1}), 
leads to a Dyson equation,  
\begin{eqnarray}
D_X({\bf q},i\Omega_n)=d_X(i\Omega_n)+d_X(i\Omega_n)M_X({\bf q},i\Omega_n)D_X({\bf q},i\Omega_n),
\end{eqnarray}
for the renormalized exciton propagator $D_X({\bf q},i\Omega_n) $ with the selfenergy 
$M_X({\bf q},i\Omega_n)$ given in Eq.~(\ref{M}). The selfenergy takes Pauli blocking into
account. Switching in Eq.~(\ref{full}) from the relative momenta ${\bf p}$ and 
${\bf p'}$ back to the original momenta ${\bf k}$ and ${\bf k'}$, we recover Eq.~(\ref{vertex}) 
for the full vertex.\\
\end{widetext}

\begin{acknowledgments}
Support from the SFB 652 is greatly acknowledged. We thank B. Bucher, 
D. Ihle, G. R\"opke, P. Wachter, and R. Zimmermann for valuable discussions
and critical reading of the manuscript.
\end{acknowledgments}


\end{document}